\documentclass[11pt]{article}

\usepackage{amssymb}
\usepackage{graphicx}
\usepackage{enumitem}
\usepackage{float}
\usepackage{chet}

\newcommand{\veps}{\varepsilon}
\newcommand{\lsp}{\hspace{1pt}}

\renewcommand{\ge}{\geqslant}

\renewcommand{\geq}{\geqslant}

\definecolor{darkblue}{rgb}{0.1,0.1,0.7}
\hypersetup{colorlinks,
           linkcolor={darkblue},
           citecolor={darkblue},
           urlcolor={darkblue}
}

\setlist{noitemsep}

\allowdisplaybreaks

\date{November 2019}

\preprint{LA-UR-19-30978}

\title{Bootstrapping Mixed Correlators in\\\vspace{5pt}
Three-Dimensional Cubic Theories II}

\author{Stefanos R.\ Kousvos$^{a,b}$ and Andreas Stergiou$^{c}$}

\affiliation{$^a$Department of Physics, University of Crete, Heraklion
GR-70013, Greece\\
$^b$Institute of Theoretical and Computational Physics (ITCP),
Department of Physics,\\\vspace{-3pt}
University of Crete, 70013 Heraklion, Greece\\
$^c$Theoretical Division, MS B285, Los Alamos National Laboratory, Los
Alamos, NM 87545, USA}

\abstract{Conformal field theories (CFTs) with cubic global symmetry in 3D
are relevant in a variety of condensed matter systems and have been studied
extensively with the use of perturbative methods like the $\varepsilon$
expansion. In an earlier work, we used the nonperturbative numerical
conformal bootstrap to provide evidence for the existence of a previously
unknown 3D CFT with cubic symmetry, dubbed ``Platonic CFT''. In this work,
we make further use of the numerical conformal bootstrap to perform a
three-dimensional scan in the space of scaling dimensions of three
low-lying operators. We find a three-dimensional isolated allowed region in
parameter space, which includes both the 3D (decoupled) Ising model and the
Platonic CFT. The essential assumptions on the spectrum of operators used
to provide the isolated allowed region include the existence of a
stress-energy tensor and the irrelevance of certain operators (in the
renormalization group sense).}

\begin{document}

\maketitle

\toc

\newsec{Introduction}
Cubic scalar theories possess global symmetry described by the 48-element
group $C_3=\mathbb{Z}_2{\!}^3\rtimes S_3\simeq S_4\times\mathbb{Z}_2\subset
O(3)$, where $S_n$ is the permutation group of $n$ objects. Their physical
interest in 3D ($d=3$ Euclidean dimensions) stems from their applications
to finite temperature magnetic and structural phase transitions
\cite{Cowley, Bruce, CowShap, Landau:1980mil, Pelissetto:2000ek}. There
exist a plethora of methods for studying them, e.g.\ the $\veps$ expansion
\cite{Pelissetto:2000ek, Adzhemyan:2019gvv, Aharony:1973zz, Osborn:2017ucf,
Rychkov:2018vya, Zinati:2019gct, Antipin:2019vdg}, the exact
renormalization group \cite{Tissier:2002zz}, Monte Carlo simulations
\cite{Caselle:1997gf} and, more recently, the numerical conformal bootstrap
\cite{Rong:2017cow, Stergiou:2018gjj, Kousvos:2018rhl}. Most studies until
recently were performed with the $\veps$ expansion, which provided up to
six-loop resummed estimates for the experimentally observable critical
exponents $\beta$ and $\nu$~\cite{Adzhemyan:2019gvv}.  Interestingly, the
critical exponents of the $\veps$ expansion, whose numerical value is
almost degenerate with those of the $O(3)$ model, agree with experiments
for magnetic phase transitions but are in tension with experiments for
structural phase transitions~\cite{RISTE19711455, PhysRevB.7.1052,
PhysRevLett.28.503, PhysRevLett.26.13, CowShap}, which appear to be closer
to Ising rather than $O(3)$ exponents.\foot{For extensive discussions on
this matter see \cite{Stergiou:2018gjj, Kousvos:2018rhl} and references
therein, and for a proposed resolution within the perturbative
renormalization group framework see \cite{PhysRevLett.33.427}.}

The study of scalar field theories with cubic symmetry via the numerical
conformal bootstrap has a short history, although the bootstrap method is
well-suited to the study of such theories.\foot{We note that analytic
bootstrap approaches in $d=4-\veps$ dimensions, such as the Mellin
bootstrap \cite{Dey:2016mcs} and the large-spin bootstrap
\cite{Alday:2017zzv, Henriksson:2018myn}, have produced critical exponents
identical to those of the ordinary diagrammatic $\veps$ expansion.} The
core constraints of any numerical bootstrap study are crossing symmetry,
i.e.\ the statement of associativity of the operator product expansion
(OPE), and unitarity.\foot{See \cite{Chester:2019wfx} for an introduction
and \cite{Poland:2018epd} for a review.} Imposing these constraints, one
may find bounds on allowed values of scaling dimensions of operators as
well as coefficients in the OPE. In certain circumstances, these bounds are
strong enough to confine scaling dimensions into isolated allowed regions
in parameter space (islands), thus providing a calculation of said scaling
dimensions which are directly linked to critical exponents. This strategy
has been previously applied with great success to $O(N)$ symmetric CFTs.
More specifically, in the case of $O(1)$ (i.e.\ the Ising model) it
provided the most precise calculation of the critical exponents to date
\cite{ElShowk:2012ht, El-Showk:2014dwa, Kos:2015mba, Kos:2014bka,
Kos:2013tga}. Extending this line of work to cubic-symmetric theories leads
to the prediction of a new fixed point \cite{Stergiou:2018gjj,
Kousvos:2018rhl}, corresponding to the theory we will call ``Platonic
CFT''. As explained in detail in~\cite{Stergiou:2018gjj, Kousvos:2018rhl},
this CFT may be relevant for cubic magnets and the structural phase
transition of SrTiO$_3$. It may also be relevant for the phase transitions
discussed in \cite{BENHASSINE2017102}, whose critical exponents
$\beta=0.306(2), \gamma=1.185(13), \delta=4.857(30)$ for LEPMO and
$\beta=0.312(11), \gamma=1.177(17), \delta=4.776(70)$ for LYPMO agree very
well with our determinations $\beta=0.308(2), \gamma=1.167(8),
\delta=4.792(11)$ for the Platonic CFT~\cite{Kousvos:2018rhl}, although it
should be noted that, according to \cite{BENHASSINE2017102}, the crystals
LEPMO and LYPMO they used crystallized in the orthorombic and not the cubic
crystal system.

\begin{figure}[ht]
  \centering \includegraphics[scale=1]{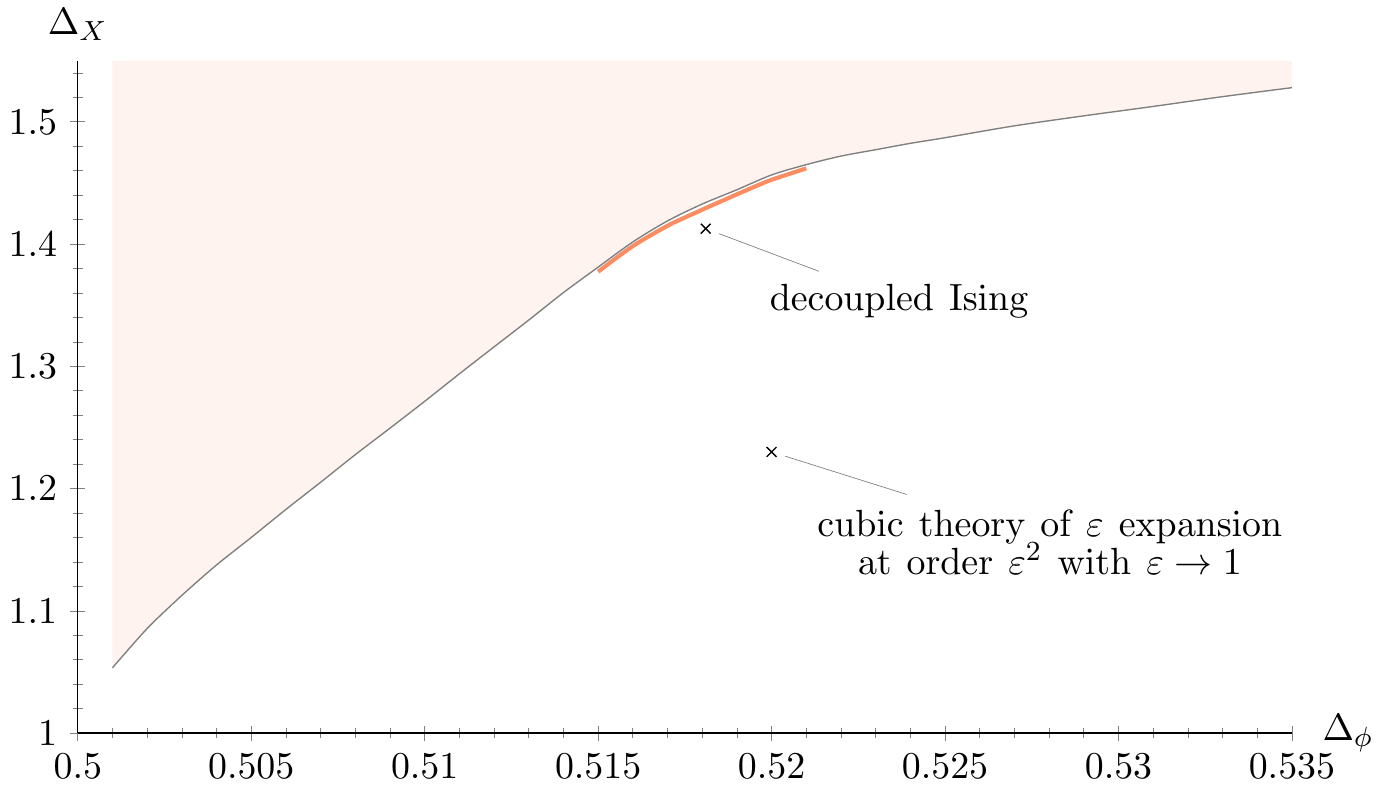}
  \caption{Upper bound on the dimension of the leading scalar $X$ operator
  in the $\phi_i\times\phi_j$ OPE. The red area is excluded. The gray bound
  is obtained using \texttt{PyCFTBoot} \cite{Behan:2016dtz} with
  $\texttt{mmax=6}$, $\texttt{nmax=9}$, $\texttt{kmax=36}$, and
  $\texttt{lmax=26}$, while the for the red bound we use stronger numerics
  with $\texttt{mmax=10}$, $\texttt{nmax=13}$, $\texttt{kmax=50}$, and
  $\texttt{lmax=40}$. The decoupled Ising theory as well as the cubic
  theory of the $\veps$ expansion are also indicated in the allowed
  region.} \label{fig:X_bound}
\end{figure}

In our previous work we performed a mixed-correlator bootstrap analysis of
four-point functions involving the leading scalar operators $\phi_i$ and
$X_{ij}$~\cite{Kousvos:2018rhl}, where $\phi_i$ transforms as a vector and
$X_{ij}$ as another nontrivial irreducible representation of the cubic
group (see below).  Our essential assumption was that the scaling dimension
of $X$ saturated the bound that was obtained for that operator dimension
using the single-correlator $\langle\phi\phi\phi\phi\rangle$ bootstrap
in~\cite{Stergiou:2018gjj}. (We reproduce that bound in
Fig.~\ref{fig:X_bound} for the reader's convenience.) The result of
\cite{Kousvos:2018rhl} was an island in the $\Delta_\phi$-$\Delta_S$ plane,
where $S$ is the leading scalar singlet of the theory, obtained by further
assuming that the next-to-leading scalar $S'$ and $X'$ operators are
irrelevant (for $S'$ the assumption was $\Delta_{S'}\ge 3.7$).

In this work we first study the effect of assumptions on operators that are
exchanged in the $\phi_i\times\phi_j$ OPE, where $\phi_i$ is the order
parameter field which is a scalar under spatial rotations and transforms
under the three-dimensional vector representation of the cubic group. In
this part we still assume that $\Delta_X$ saturates the bound of Fig.\
\ref{fig:X_bound}.  The $\phi_i\times\phi_j$ OPE can be written as
\eqn{\phi_i\times\phi_j\sim \delta_{ij}\lsp S + X_{(ij)} + Y_{(ij)} +
A_{[ij]}\,,}[phiphiOPE]
where $S$, $X_{(ij)}$ and $Y_{(ij)}$ are even-spin operators and $
A_{[ij]}$ are odd-spin operators. If the cubic group is viewed as a
subgroup of $O(3)$, $X$ and $Y$ form respectively the diagonal and
non-diagonal parts of the two-index traceless-symmetric representation of
$O(3)$ (conveniently thought of as a $3\times 3$ matrix). Due to the fact
that the cubic group is discrete, there is no global-symmetry conserved
current in $A_{[ij]}$. The leading spin-two singlet in any local CFT is the
stress-energy tensor, whose dimension is fixed to be equal to $d$ as a
result of conservation.  Below we will examine the effects of assuming that
there is a stress-energy tensor $T_{\mu\nu}$ and a gap in the scaling
dimension of $T_{\mu\nu}$ and the next-to-leading spin-two singlet,
$T_{\mu\nu}^\prime$.

The primary aim of this work is to eliminate the assumption that the
leading scalar $X$ operator lies on the bound of Fig.~\ref{fig:X_bound}. To
that end, we study unitarity and crossing symmetry constraints on a system
of correlators which consists of $\langle\phi\phi\phi\phi\rangle$,
$\langle\phi\phi SS\rangle$ and $\langle SSSS\rangle$, where $S$ is the
scalar singlet with the lowest scaling dimension (besides the obligatory
unit operator) that appears in the OPE of $\phi_i$ with itself. Still
assuming the existence of $T_{\mu\nu}$ and a gap to the dimension of
$T_{\mu\nu}^\prime$, along with further assumptions on some specific
operators listed below, we are able to find a three-dimensional convex
island that includes the decoupled Ising and the Platonic CFT. Within the
set of assumptions we are using, we are unable to find two separate
islands, one for each theory.

The structure of the present work is as follows. In section \ref{secA} we
analyze the required group theory and obtain the constraints resulting from
a single correlator system. In section \ref{secB} we analyze the system of
multiple correlators involving $\phi$ and $S$, and derive numerically the
three-dimensional island.

\newsec{Single correlator}[secA]
\subsec{OPE, four-point function, and crossing equation}
In order to work out the required tensor structures of the global symmetry
for the four-point function, it is rather convenient to work with the cubic
group as a subgroup of $O(3)$. Schematically, the OPE of an $O(3)$ vector
with itself takes the form
\begin{equation}
  \phi_i  \times \phi_j \sim \delta_{ij} S + T_{(ij)}+A_{[ij]}\,,
\label{1}
\end{equation}
where $S$ is the singlet, $T$ the traceless-symmetric and $A$ the
antisymmetric representation. We note that in the cubic case the
traceless-symmetric representation splits into its diagonal and
non-diagonal parts; each one furnishes an irreducible representation
(irrep) of the cubic group. Thus we have
\begin{equation}
  \phi_i  \times \phi_j \sim \delta_{ij} S + X_{(ij)}+ Y_{(ij)}+A_{[ij]}\,,
\label{2}
\end{equation}
where $X$ corresponds to the diagonal irrep and $Y$ to the non-diagonal
irrep. The CFT of three decoupled Ising models has cubic symmetry. In that
case the $X$ operators are given by sums of operators from each Ising
model, while the $Y$ operators involve sums of products of operators from
each Ising model. The lowest-dimension scalar $X$ and $Y$ operators in the
decoupled Ising theory have scaling dimensions
$\Delta_X=\Delta_\epsilon\approx 1.4126$ and
$\Delta_Y=2\Delta_\sigma\approx 1.0362$, respectively.

We wish to study the four-point function $ \langle \phi_i \phi_j \phi_k
\phi_l \rangle$. In order to decompose it onto irreps of the cubic group we
use \eqref{2} and get, in the $12\to34$ channel,
\begin{equation}
\langle \phi_i \phi_j \phi_k \phi_l \rangle \sim
\delta_{ij}\delta_{kl}\langle SS\rangle+\langle X_{ij}X_{kl}\rangle+\langle
Y_{ij}Y_{kl}\rangle+\langle A_{ij}A_{kl}\rangle\,.
\label{3}
\end{equation}
Noticing that that the tensor structures of the $X$ and $Y$ representations
must add up to the one of the $T$ representation of $O(3)$, and demanding
that they be orthogonal to each other as well as diagonal/non-diagonal,
respectively, we obtain the following expressions for the global symmetry
projectors of the four-point function:
\begin{equation}
\begin{gathered}
P^S_{ijkl}=\tfrac{1}{3}\delta_{ij}\delta_{kl}\,,\qquad
P^X_{ijkl}=\delta_{ijkl}-\tfrac{1}{3}\delta_{ij}\delta_{kl}\,,
\\
P^Y_{ijkl}=-\delta_{ijkl}+\tfrac{1}{2}(\delta_{ik}\delta_{jl}+\delta_{il}\delta_{jk})\,,\qquad
P^A_{ijkl}=\tfrac{1}{2}(\delta_{il}\delta_{jk}-\delta_{ik}\delta_{jl})\,,
\end{gathered}
\label{4}
\end{equation}
where the numeric prefactor of each projector is chosen in such a way that they are orthonormal to each other. It can easily be checked that $ P^X_{ijkl}+P^Y_{ijkl}=P^T_{ijkl}$. With this in hand we have everything required to derive the crossing equations.

The crossing equation follows from demanding that the $12\to34$
decomposition of the four-point function is equal to the $14\to32$ one.  We
identify four equations, corresponding to terms multiplying
$\delta_{ij}\delta_{kl}$, $\delta_{il}\delta_{jk}$,
$\delta_{ik}\delta_{jl}$ and $\delta_{ijkl}$ which must independently be
equal to zero. Hence, we arrive at our system of crossing equation sum
rules:
\begin{align}
\begin{split}
  \sum_{Y^+} \lambda^2_{O_Y} F^-_{\Delta,\ell}+\sum_{A^-} \lambda^2_{O_A}
  F^-_{\Delta,\ell}=0\,,
  \\
  \tfrac{1}{3}\sum_{S^+} \lambda^2_{O_S}
  F^-_{\Delta,\ell}-\tfrac{2}{3}\sum_{X^+} \lambda^2_{O_X}
  F^-_{\Delta,\ell}+\sum_{Y^+} \lambda^2_{O_Y} F^-_{\Delta,\ell}-\sum_{A^-}
  \lambda^2_{O_A} F^-_{\Delta,\ell}=0\,,
  \\
  \tfrac{1}{3}\sum_{S^+} \lambda^2_{O_S}
  F^+_{\Delta,\ell}-\tfrac{2}{3}\sum_{X^+} \lambda^2_{O_X}
  F^+_{\Delta,\ell}-\sum_{Y^+} \lambda^2_{O_Y} F^+_{\Delta,\ell}+\sum_{A^-}
  \lambda^2_{O_A} F^+_{\Delta,\ell}=0\,,
  \\
  \tfrac{1}{3}\sum_{S^+} \lambda^2_{O_S}
  F^-_{\Delta,\ell}+\tfrac{4}{3}\sum_{X^+} \lambda^2_{O_X}
  F^-_{\Delta,\ell}=0\,,
  \label{cr2}
\end{split}
\end{align}
where
\begin{equation}
  F^{\pm}_{\Delta,\ell}=v^{\Delta_\phi}g_{\Delta,\ell}(u,v)\pm
  u^{\Delta_\phi}g_{\Delta,\ell}(v,u)
  \label{F1}
\end{equation}
and $ g$ is the corresponding conformal block for the exchange of an
operator with scaling dimension $ \Delta$ and spin $\ell$, while $
u=\frac{x_{12}^2x_{34}^2}{x_{13}^2x_{24}^2}$ and $
v=\frac{x_{14}^2x_{23}^2}{x_{13}^2x_{24}^2}$ are the usual conformal cross
ratios and $x_{ij}=\sqrt{(x_i-x_j)^2}$. Lastly, the plus or minus
superscript on operators being summed over indicates summation over even or
odd spins, and for the OPE coefficients $\lambda_{O}$ is shorthand for
$\lambda_{\phi \phi O}$.

\subsec{Numerical results}
Before proceeding, we note that a collection of results regarding bootstrap
bounds obtained with assumptions on conserved currents for various symmetry
groups have appeared previously in \cite{Li:2017kck}.  Our main assumption
in this section is that the lowest-dimension scalar operator in the $X$
irrep saturates its corresponding bootstrap bound in
Fig.~\ref{fig:X_bound}.\foot{Here we do not use the scalar singlet bound,
for our theory of interest does not saturate that bound. It has been
observed in practice that the bootstrap of subgroups of $O(N)$, for some
value of $N$, always present scalar singlet sector bounds identical to
$O(N)$ ones. Known examples include hypercubic and hypertetrahedral
theories \cite{Stergiou:2018gjj}, MN and tetragonal theories
\cite{Stergiou:2019dcv} and $O(m)\times O(n)$ theories
\cite{Nakayama:2014sba, Nakayama:2014lva, progress}.} The reason for this
assumption is twofold. Firstly, bootstrap intuition tells us that kinks
(abrupt changes in slope) in bootstrap bounds correspond to the positions
in parameter space where physical CFTs live (which is precisely how the
Ising and $O(N)$ CFTs were first discovered in the context of the conformal
bootstrap). Such a kink is indeed observed in Fig.~\ref{fig:X_bound}.
Secondly, such an assumption is needed to obstruct the enhancement of the
symmetry from cubic to $O(3)$. Such an enhancement requires
$\Delta_X=\Delta_Y$, something that cannot happen because the bound on
$\Delta_Y$ is much below the bound on $\Delta_X$ \cite[Fig.\
3]{Stergiou:2018gjj}.

The points used to saturate the $X$ sector bound are calculated with a
vertical precision of $10^{-9}$. To produce the required \texttt{xml} files
we used \texttt{PyCFTBoot} \cite{Behan:2016dtz} with the following
parameters: $\texttt{mmax=6}$, $\texttt{nmax=9}$, $\texttt{kmax=36}$, and
$\texttt{lmax=26}$. All the plots presented in this section have a vertical
precision of $10^{-6}$. For the optimization of the bootstrap problem we
use \texttt{SDPB}~\cite{Simmons-Duffin:2015qma, Landry:2019qug}.

\begin{figure}[H]
  \centering
  \includegraphics[scale=1]{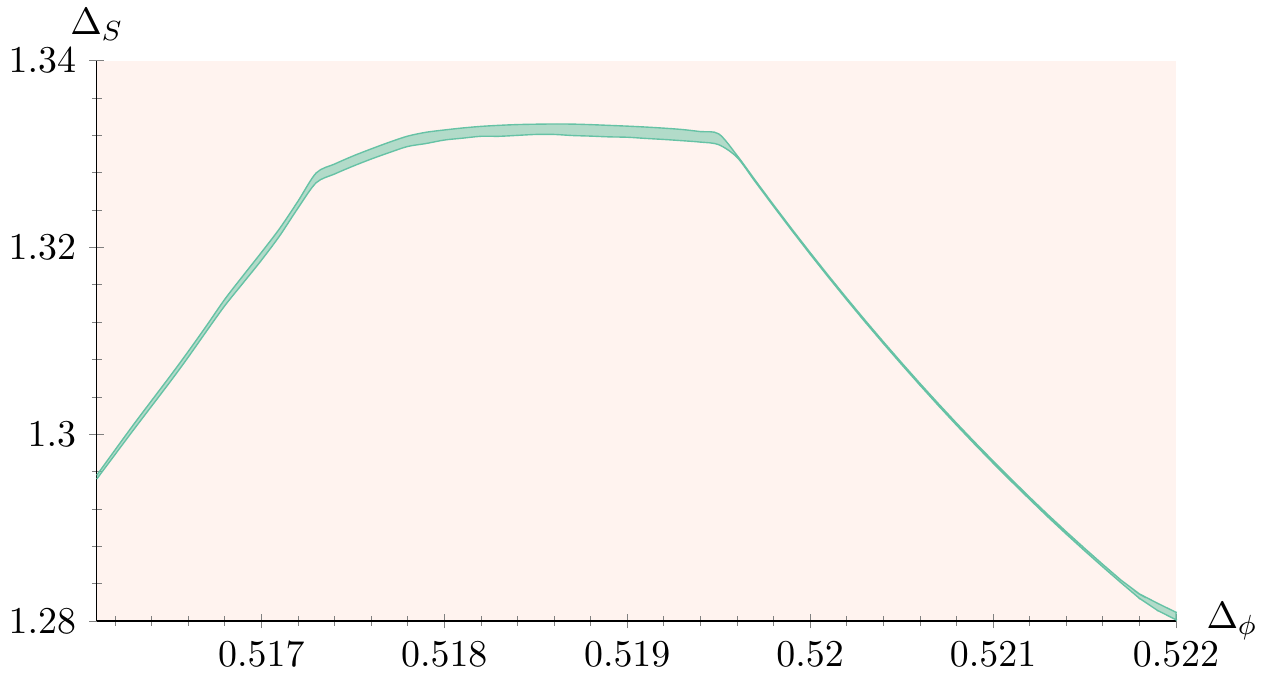}
  \caption{Plot of the allowed region (in green) with the assumptions
  \ref{assumIitemI}--\ref{assumIitemIV}.  There appear to be two kinks in
  the allowed strip. It will become apparent from Fig.\ \ref{fig2} that
  these two kinks arise at the points where this plot overlaps with the
  peninsula of allowed parameter space found in our previous
  work~\cite{Kousvos:2018rhl}.}
  \label{fig1}
\end{figure}

First we obtain a plot, Fig.~\ref{fig1}, using the following assumptions:
\begin{enumerate}[label=(S-\arabic*), leftmargin=*]
  \item saturation of $X$ bound of
    Fig.~\ref{fig:X_bound},\label{assumIitemI}
  \item existence of stress-energy tensor $T_{\mu\nu}$, i.e.\ $\Delta_{T_{\mu\nu}}=3$,
   \label{assumIitemII}
  \item dimension of next-to-leading spin-two singlet, $T_{\mu\nu}^\prime$,
    above 4, i.e.\ $\Delta_{T_{\mu\nu}^\prime}\ge4$,
    \label{assumIitemIII}
  \item dimension of next-to-leading scalar singlet, $S'$, above
    3, i.e.\ $\Delta_{S'}\geq3$.\label{assumIitemIV}
\end{enumerate}
Note that the allowed region does not yet truncate on either side.
Assumptions (S-3) and (S-4) are motivated by the extremal functional
method~\cite{El-Showk:2016mxr} (for (S-4) see
\cite[Fig.~7]{Stergiou:2018gjj}).

\begin{figure}[H]
  \centering
  \includegraphics{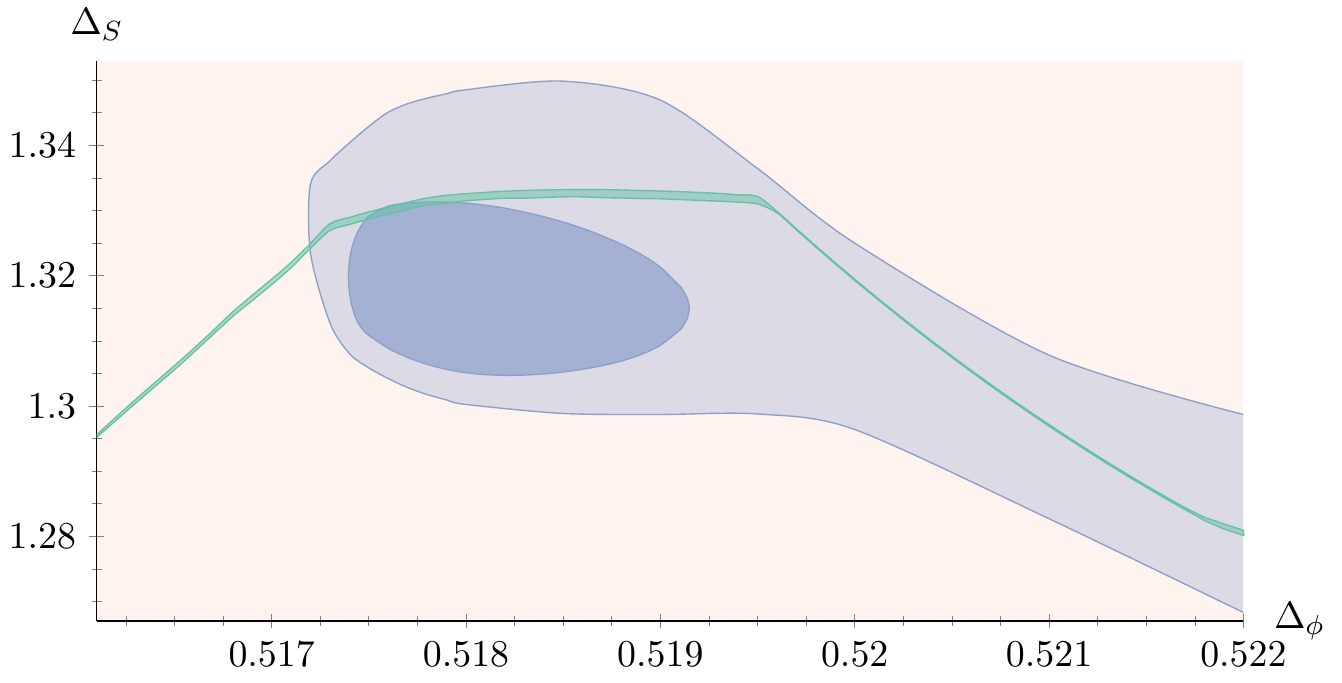}
  \caption{In this figure we observe the overlap between the
  single-correlator allowed region from this work (in green), and the
  allowed peninsula and island from our previous mixed-correlator work
  \cite{Kousvos:2018rhl} (in blue). Both the peninsula and island assume
  that the theory lives on the $X$ sector single-correlator bound of
  Fig.~\ref{fig:X_bound}, and also that the next-to-leading operator in the
  scalar $X$ sector has a scaling dimension of $3$ or higher. Additionally,
  the island assumes that $S'$ has a dimension of $3.7$ or higher. The
  peninsula assumes that $S'$ has a dimension of $3$ or higher.}
  \label{fig2}
\end{figure}

We may compare this to the allowed region studied in our previous
mixed-correlator work (which did not assume the existence of a conserved
stress-energy tensor); this is done in Fig.\ \ref{fig2}. Lastly, we may
look at how the allowed region changes when also imposing a gap between the
$X$

\begin{figure}[H]
  \centering
  \includegraphics{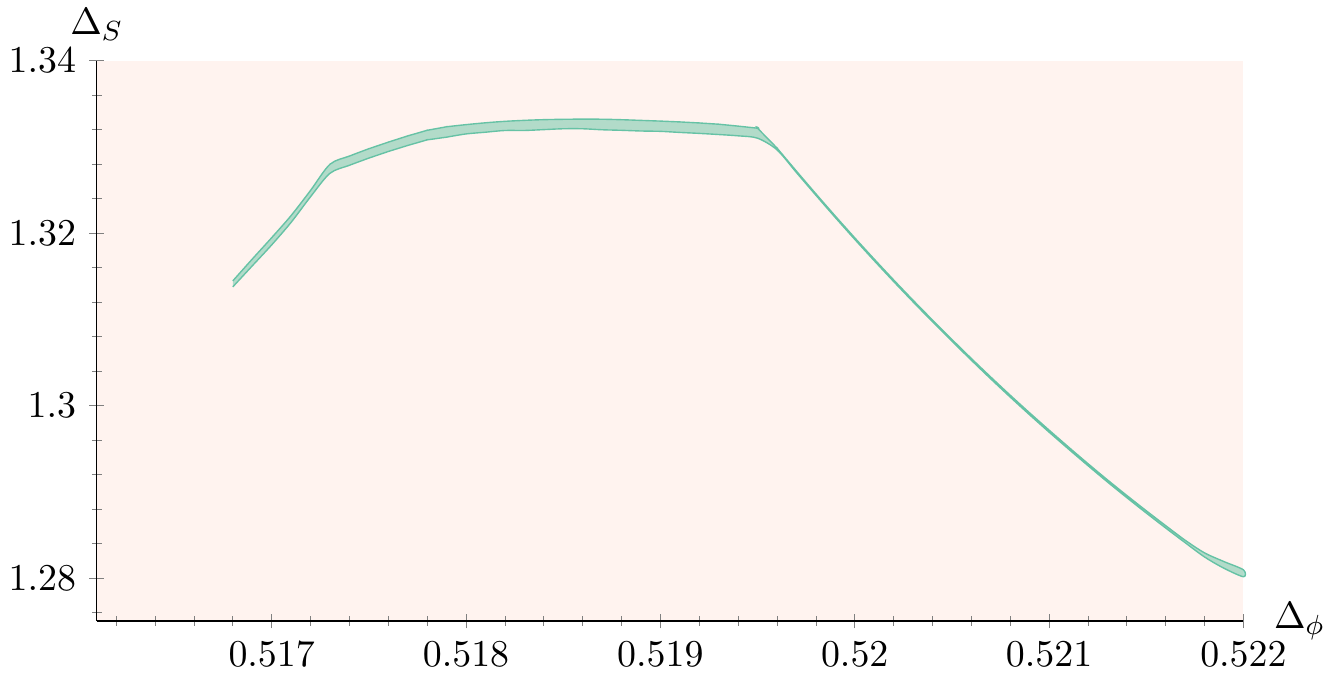}
  \caption{In this plot we assume, in addition to
  \ref{assumIitemI}--\ref{assumIitemIV}, that $X'$ has a dimension of 2.8
  or higher. This extra assumption leads to the truncation of the allowed
  region on the left.}
  \label{fig3}
\end{figure}

\noindent operator saturating the bootstrap bound of Fig.\
\ref{fig:X_bound} and the next operator in the same irrep, $X'$---this is
done in Fig.\ \ref{fig3}.  There we see that the effect of raising the gap
on $X'$ is that the allowed strip truncates on the left-hand side but
remains unaffected on the right-hand side (compared to the strip in
Fig.~\ref{fig1}). We have checked that changing the gap on $S'$ between $3$
and $3.7$ leaves Fig.~\ref{fig3} essentially unchanged.

These gaps do not produce an island as we would ideally desire. In the next
section we will see that, using a mixed-correlator system, we will be able
to relax the assumptions on the $X$ irrep (i.e.\ we will no longer demand
that it lives on the single correlator bound) and obtain an isolated
allowed region in parameter space.

\newsec{Multiple correlators}[secB]
\subsec{OPEs, four-point functions, and crossing equations}
In this section we study the additional correlators needed for the three
dimensional scan of parameter space.  First, we must note that in this work
we consider a different system of mixed correlators compared to our
previous work \cite{Kousvos:2018rhl}.  This brings considerable
simplifications in terms of the group theory structures required. We need
the OPE of the vector operator with the scalar singlet $S$, and the OPE of
the scalar singlet $S$ with itself. These follow trivially:
\begin{equation}
\phi_i \times S \sim \phi_i\,,
\label{5}
\end{equation}
and
\begin{equation}
  S \times S\sim S\,.
\label{6}
\end{equation}

Using \eqref{2}, \eqref{5} and \eqref{6} we obtain
\begin{equation}
\langle \phi_i S \phi_j S \rangle \sim \langle \phi_i \phi_j
\rangle\,,\qquad
\langle \phi_i \phi_j S S \rangle \sim \delta_{ij}\langle S S \rangle\,,
\label{7}
\end{equation}
where we take the OPE of the first two and last two operators in the four-point function.

The crossing equations are
\begin{align}
\begin{split}
\sum_{S^+}\lambda^2_{SSO_S}F^{-\lsp SS,SS}_{\Delta,\ell}=0\,,
\\
\sum_{V^\pm}\lambda^2_{\phi SO_V}F^{-\lsp\phi S,\phi S}_{\Delta,\ell}=0\,,
\\
\sum_{S^+}\lambda_{\phi \phi O}\lambda_{S S O_S}
F^{\mp\lsp \phi\phi,S S}_{\Delta,\ell}\pm
\sum_{V^\pm}(-1)^\ell\lambda^2_{\phi S O_V}
F^{\mp\lsp S\phi , \phi S}_{\Delta,\ell}=0\,,
\label{cr4}
\end{split}
\end{align}
where again the plus or minus superscripts on operators are used to
indicate summation over even or odd spin operators, respectively, and
\begin{equation}
F^{\pm\lsp
ij,kl}_{\Delta,\ell}=v^{\frac{\Delta_j+\Delta_k}{2}}g_{\Delta,\ell}^{ij,kl}
(u,v)\pm
u^{\frac{\Delta_j+\Delta_k}{2}}g_{\Delta,\ell}^{ij,kl}(v,u)\,.
\label{F2}
\end{equation}
The numerical bootstrap problem can now be set up along the lines described
in~\cite{Kos:2014bka}.

\subsec{Numerical results}
Equipped with the machinery of mixed correlators, we may now obtain a
closed isolated region in parameter space. This works as follows. It is
important, just as in the previous section, to obstruct the enhancement of
the symmetry to $O(3)$. This is done by assuming that the scaling dimension
of the first $X$ operator is equal to, or greater than $1.4126$. This is in
the allowed region of Fig.~\ref{fig:X_bound} for $\Delta_\phi\gtrsim
0.5165$, and it obstructs the enhancement of the symmetry to $O(3)$ due to
the fact that $\Delta_X \geq 1.4126$ lies in the disallowed region for the
scaling dimension of the leading scalar $Y$ sector operator for
$\Delta_\phi$ in our region of interest (see \cite[Fig.\
3]{Stergiou:2018gjj}).\foot{In the $O(3)$ theory $X$ and $Y$ have the same
scaling dimension, for they combine to form the two-index traceless
symmetric of $O(3)$.} Note that in \cite{Stergiou:2018gjj} it was observed
that all $X$ sector hypercubic kinks lie above the Ising model, which they
approach from above as $N\to\infty$. (This also motivates our choice of the
Ising gap in assumption \ref{assumIIitemI} below.)

\begin{figure}[ht]
  \centering
  \includegraphics[scale=1]{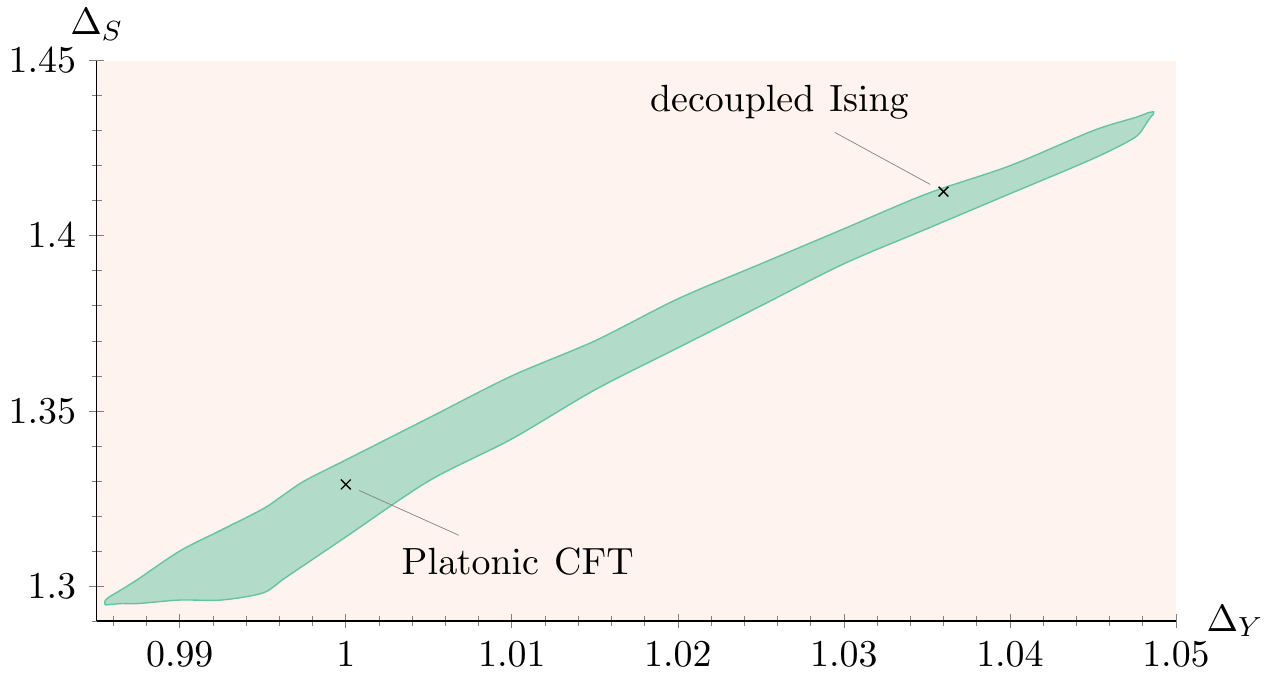}
  \caption{Plot of the projection of the three-dimensional allowed region
  in ($\Delta_\phi$,$\Delta_S$,$\Delta_Y$) onto the $\Delta_S$-$\Delta_Y$
  plane using the assumptions \ref{assumIIitemI}--\ref{assumIIitemVI}.}
  \label{fig4}
\end{figure}

\begin{figure}[ht]
  \centering
  \includegraphics[scale=1]{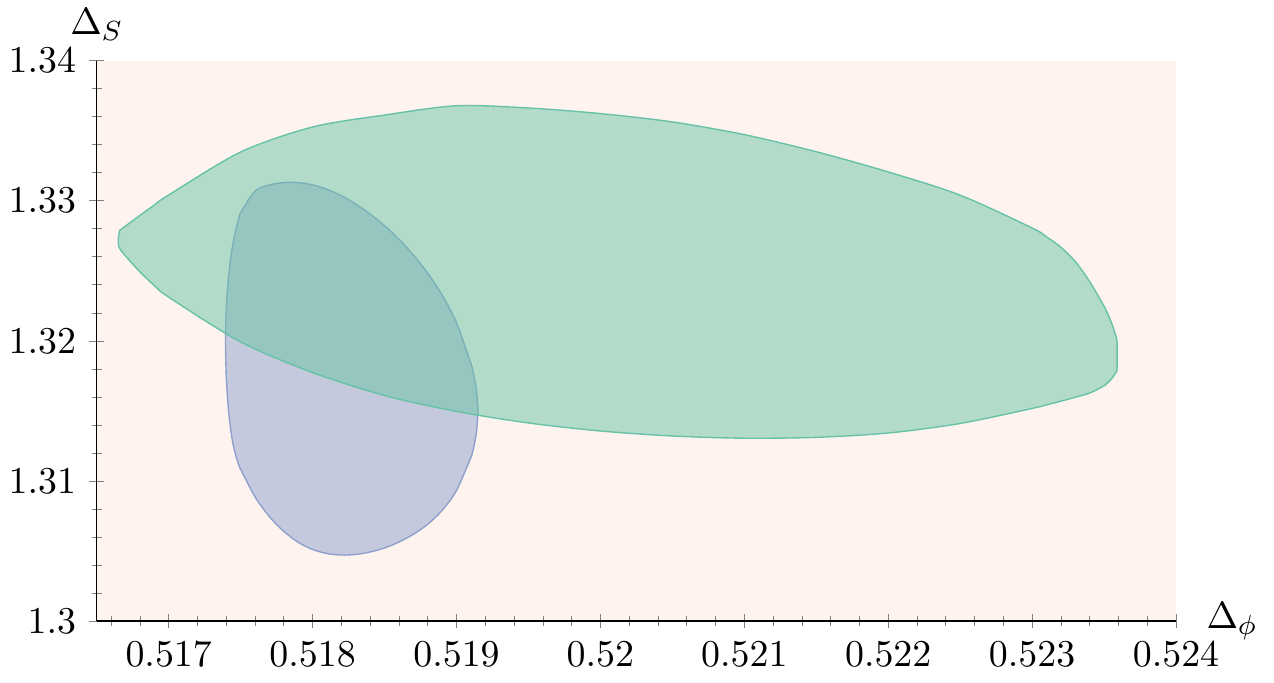}
  \caption{Plot of the allowed region in the $\Delta_\phi$-$\Delta_S$
  plane, derived with $\Delta_Y=1$ and using the assumptions
  \ref{assumIIitemI}--\ref{assumIIitemVI}. The blue island is that
  of~\cite[Fig.~7]{Kousvos:2018rhl} with $\Delta_{S'}\ge3.7$.}
  \label{fig5}
\end{figure}

\begin{figure}[ht]
  \centering
  \includegraphics[scale=1]{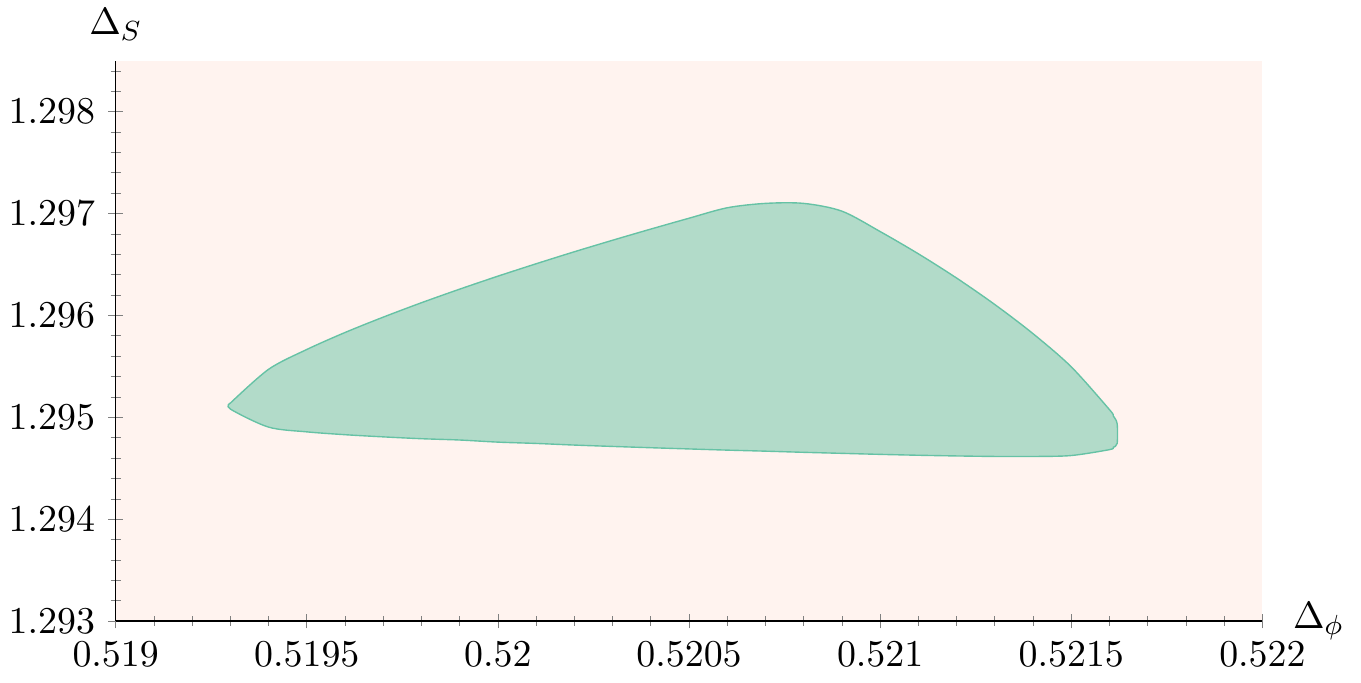}
  \caption{Plot of the allowed region in the $\Delta_\phi$-$\Delta_S$
  plane, derived with $\Delta_Y=0.98575$ and using the assumptions
  \ref{assumIIitemI}--\ref{assumIIitemVI} .}
  \label{fig6}
\end{figure}

\begin{figure}[ht]
  \centering
  \includegraphics[scale=1]{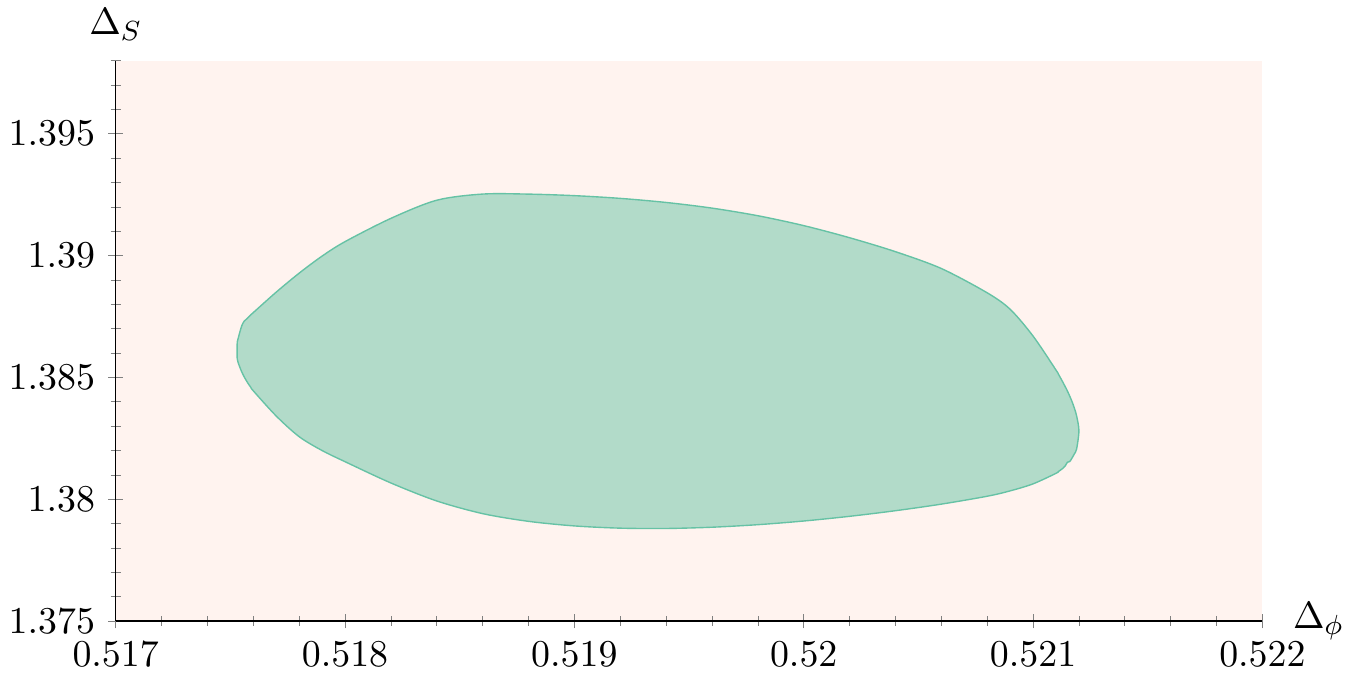}
  \caption{Plot of the allowed region in the $\Delta_\phi$-$\Delta_S$
  plane, derived with $\Delta_Y=1.025$ and using the assumptions
  \ref{assumIIitemI}--\ref{assumIIitemVI}.}
  \label{fig7}
\end{figure}

The assumptions we make are summarized below:
\begin{enumerate}[label=(M-\arabic*), leftmargin=*]
  \item $\Delta_X\ge1.4126$,\label{assumIIitemI}
  \item existence of conserved stress-energy tensor, $T_{\mu\nu}$,  i.e.\ $\Delta_{T_{\mu\nu}}=3$,
    \label{assumIIitemII}
  \item $\Delta_{T_{\mu\nu}^\prime}\ge4$,\label{assumIIitemIII}
  \item $\Delta_{S'}\ge3.7$,\label{assumIIitemIV}
  \item $\Delta_{Y'}\ge 3.0$,\label{assumIIitemV}
  \item $\Delta_{\phi'}\ge 1.5$.\label{assumIIitemVI}
\end{enumerate}
We also impose the equality of OPE coefficients $\lambda_{\phi\phi
S}=\lambda_{\phi S\phi}$. Here $Y'$ is the next-to-leading scalar operator
in the $Y$ irrep of the $\phi\times\phi$ OPE and $\phi'$ is the
next-to-leading operator in the $\phi\times S$ OPE ($\phi$ is the leading
one). The assumption (M-6) is the statement that the cubic in $\phi$
operator has dimension above the free theory dimension of $\phi^2\phi_i$.
With these assumptions we perform multiple scans: for each scan we fix a
dimension $\Delta_Y$ for the leading $Y$ operator and find the lowest and
highest admissible value for $\Delta_S$. This procedure leads to Fig.\
\ref{fig4}, which is obtained by finding the allowed region in the
three-dimensional space parametrized by
($\Delta_\phi$,$\Delta_S$,$\Delta_Y$) and consequently projecting it onto
the $\Delta_Y$-$\Delta_S$ plane.  Indicatively, we show three different
slices of this three-dimensional space, obtained by fixing $\Delta_Y=1$,
$\Delta_Y=0.98575$ and $\Delta_Y=1.025$ and looking at the
$\Delta_\phi$-$\Delta_S$ plane in Figs.\ \ref{fig5}, \ref{fig6} and
\ref{fig7}, respectively. The parameters we have used in \texttt{PyCFTBoot}
\cite{Behan:2016dtz} are, $\texttt{mmax=5}$, $\texttt{nmax=7}$,
$\texttt{kmax=36}$, and $\texttt{lmax=26}$ . Depending on the position in
Fig.\ \ref{fig4}, the vertical precision starts from $0.002$ and goes up to
$ 0.0001$ (the increased precision is used at the edges of the island for
increased smoothness).


We find that the right-hand side of the island in Fig.~\ref{fig4} truncates
close to the Ising fixed point, whereas the left-hand side of the island
truncates close to the Platonic fixed point. Note that, using the extremal
functional method \cite{El-Showk:2016mxr} on the $X$ sector single
correlator bound of Fig.~\ref{fig:X_bound}, we find a scaling dimension
$\Delta_Y$ numerically very close to $1$ for the Platonic CFT.\foot{We note
that although both decoupled Ising and Platonic CFT are denoted by a cross
in Fig.\ \ref{fig4}, the position of the Platonic CFT is approximate and
not known with the precision of the decoupled Ising CFT.} This is
consistent with the island presented here. For $\Delta_Y=1$ one may view
Fig.\ \ref{fig5}, where the large green island corresponds to the one found
in this work and the smaller blue one to that of
\cite[Fig.~7]{Kousvos:2018rhl} with $\Delta_{S'}\ge3.7$. The Platonic CFT
must lie in the overlap of the two islands of Fig.~\ref{fig5}.

As can be seen from Fig.\ \ref{fig6}, the island shrinks in size as we
approach the left-most edge of the allowed region of Fig.\ \ref{fig4}, but
without approaching $\Delta_\phi=0.5165$ (the $\Delta_\phi$ above which
\ref{assumIIitemI} becomes valid). Similarly, Fig.\ \ref{fig7} shows that
the island shrinks in size as we approach the right-most edge of the
allowed region of Fig.\ \ref{fig4}, again without approaching
$\Delta_\phi=0.5165$. Hence, our three-dimensional island is isolated
because of assumptions \ref{assumIIitemII}--\ref{assumIIitemVI}, and
assumption \ref{assumIIitemI} simply serves the purpose of precluding the
enhancement of the global symmetry to $O(3)$.

\newsec{Concluding remarks}
In this work we studied the parameter space surrounding the Platonic CFT.
This was done using a mixed-correlator bootstrap, with the external
operators being scalar (spin-zero) operators that transform in the vector
and singlet representation of the cubic group.  In our previous work we had
studied a mixed-correlator system involving the vector of the cubic group,
but instead of the singlet we had considered a scalar in the diagonal
representation we call $X$~\cite{Kousvos:2018rhl}. That work had produced
an island in the $\Delta_\phi$-$\Delta_S$ plane, but it involved the
assumption that $\Delta_X$ saturated its bootstrap bound (shown here in
Fig.~\ref{fig:X_bound}). The main motivation of this work was to remove
that assumption. Even without it, we find a three-dimensional island,
two-dimensional projections of which can be seen in Figs.\ \ref{fig4} and
\ref{fig5}. Our results further establish and solidify the existence of the
Platonic CFT.

The island we find in this work includes the Platonic and the decoupled
Ising CFT (see Fig.~\ref{fig4}).\foot{We remind the reader that they both
have the same global symmetry, namely cubic.} We were not able to separate
the allowed region into two islands, one around each CFT. This demonstrates
that the low-lying spectrum of the two theories is very similar, at least
in the sectors we have examined, although their critical exponents are
manifestly different.  The crucial spectrum assumption that can be used to
distinguish these two CFTs and allow us to obtain two distinct islands
around them remains unknown.

An important result established here is that the Platonic CFT contains a
conserved stress-energy tensor, something that implies its locality. This
result is nontrivial due to the fact that currently we do not have a
microscopic understanding of the Platonic CFT, i.e.\ we are not aware of a
Lagrangian that flows to it, as is the case, for example, for the Ising
model in $d=3$.

This work offers independent constraints on operators that have been
already constrained using other channels in our earlier work. More
specifically, scalar singlets appear in the $\phi\times\phi$ and $X\times
X$ OPEs, see \cite{Stergiou:2018gjj} and \cite{Kousvos:2018rhl},
respectively, while they also appear in the $S\times S$ OPE we considered
in this work. The overall consistency of our results is corroborated by our
results here, which suggest that the dimension of the next-to-leading
scalar singlet is well above marginality, specifically $\Delta_{S'}\approx
3.7$. This means that the Platonic CFT has only one relevant scalar singlet
operator, and thus it corresponds to a critical theory in the usual
classification.

This work also further supports the conclusion that the Platonic CFT is not
the cubic theory found with the $\veps$ expansion in $d=4-\veps$. Indeed,
the critical exponents determined for the cubic theory of the $\veps$
expansion~\cite{Adzhemyan:2019gvv} are very different from those determined
from our results in this work (using the overlap in Fig.~\ref{fig5} for
example), which are consistent with our earlier results
in~\cite{Kousvos:2018rhl}.

An interesting future direction would be to perform a mixed-correlator
bootstrap with $\phi$, $X$ and $S$ as external operators, and identify the
assumptions with which we may obtain a three-dimensional island in the
$(\Delta_\phi, \Delta_X, \Delta_S)$ space.  It would also be of great
interest to use the bootstrap to study the predictions of the $\veps$
expansion for the cubic theory, i.e.\ use spectrum assumptions based on
$\veps$ expansion results in order to move into the allowed region of
Fig.~\ref{fig:X_bound} and nonperturbatively analyze the cubic theory
predicted by the $\veps$ expansion.


\ack{SRK would like to thank Slava Rychkov for his hospitality and support
during his extended stay at the Institut des Hautes \'Etudes Scientifiques
(IHES) where part of this work was completed; this stay was supported by
the Simons Foundation grant 488655 (Simons Collaboration on the
Nonperturbative Bootstrap) and by Mitsubishi Heavy Industries via an
ENS-MHI Chair. SRK would also like to thank the IHES for their hospitality
during his extended stay there. The research work of SRK is supported by
the Hellenic Foundation for Research and Innovation (HFRI) under the  HFRI
PhD Fellowship grant (Fellowship Number: 1026). Research presented in this
article was supported by the Laboratory Directed Research and Development
program of Los Alamos National Laboratory under project number
20180709PRD1. The numerical computations in this paper were run on the
LXPLUS cluster at CERN and the Metropolis cluster at the Crete Center for
Quantum Complexity and Nanotechnology.
\vspace{-0.8cm}
\begin{figure}[H]
  \flushright
  \includegraphics[scale=0.45]{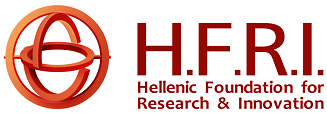}
\end{figure}
}
\vspace{-1.8cm}
\newpage
\bibliography{bootstrapping_mixed_cubic_correlators}

\end{document}